\documentclass[english,prd,superscriptaddress,nofootinbib,preprintnumbers]{revtex4}
\usepackage[T1]{fontenc}
\usepackage[latin1]{inputenc}
\usepackage{amsmath}
\usepackage{amssymb}
\usepackage{esint}

\makeatletter
\@ifundefined{textcolor}{}
{%
 \definecolor{BLACK}{gray}{0}
 \definecolor{WHITE}{gray}{1}
 \definecolor{RED}{rgb}{1,0,0}
 \definecolor{GREEN}{rgb}{0,1,0}
 \definecolor{BLUE}{rgb}{0,0,1}
 \definecolor{CYAN}{cmyk}{1,0,0,0}
 \definecolor{MAGENTA}{cmyk}{0,1,0,0}
 \definecolor{YELLOW}{cmyk}{0,0,1,0}
}

\usepackage{babel}

\makeatother

\usepackage{babel}
\begin{document}

\title{Vainshtein mechanism in second-order scalar-tensor theories}

\author{Antonio De Felice}
\affiliation{TPTP \& NEP, The Institute for Fundamental Study, Naresuan University,
Phitsanulok 65000, Thailand}
\affiliation{Thailand Center of Excellence in Physics, Ministry of Education,
Bangkok 10400, Thailand}

\author{Ryotaro Kase}
\affiliation{Department of Physics, Faculty of Science, Tokyo University of Science,
1-3, Kagurazaka, Shinjuku-ku, Tokyo 162-8601, Japan}

\author{Shinji Tsujikawa}
\affiliation{Department of Physics, Faculty of Science, Tokyo University of Science,
1-3, Kagurazaka, Shinjuku-ku, Tokyo 162-8601, Japan}

\begin{abstract}
In second-order scalar-tensor theories we study how the Vainshtein
mechanism works in a spherically symmetric background with a matter
source. In the presence of the field coupling $F(\phi)=e^{-2Q\phi}$
with the Ricci scalar $R$ we generally derive the Vainshtein radius
within which the General Relativistic behavior is recovered even for
the coupling $Q$ of the order of unity. Our analysis covers the models
such as the extended Galileon and Brans-Dicke theories with a dilatonic
field self-interaction. We show that, if these models are responsible
for the cosmic acceleration today, the corrections to gravitational
potentials are generally small enough to be compatible with local
gravity constraints.
\end{abstract}

\date{\today}

\maketitle

\section{Introduction}

The observational discovery of the late-time cosmic acceleration \cite{snia}
poses one of the most serious problems in modern cosmology. Within
the framework of General Relativity (GR) it is possible to realize
the accelerated expansion of the Universe by taking into account a
``dark'' component of a matter source. The representative model
of this class is quintessence, in which the source of dark energy
comes from the potential energy of a scalar field $\phi$ \cite{quin}.
However, it is generally difficult to accommodate an extremely tiny
mass required for the cosmic acceleration today ($m_{\phi}\approx10^{-33}$\,eV)
in the framework of particle physics \cite{Kolda}.

An alternative approach to the dark energy problem is the modification
of gravity at large distances \cite{review}. One of the simplest
examples is the so-called $f(R)$ gravity, in which the Lagrangian
$f$ is a general function of the Ricci scalar $R$ \cite{fR}. The
functions $f(R)$ are required to be carefully designed to satisfy
cosmological and local gravity constraints \cite{Hu07,Star07,Appleby,Tsuji07,Linder}.
In viable dark energy models based on $f(R)$ gravity the mass of
a scalar degree of freedom is large in high-density regions, so that
the chameleon mechanism \cite{chame} can be at work to suppress the
propagation of the fifth force. Nevertheless, in the cosmological
context, such models are plagued by the fine tuning of initial conditions
associated with the oscillating mode of field perturbations \cite{Star07,Tsuji07,Appleby2,Frolov}.
This property generally persists for the chameleon models of dark
energy which are designed to pass both cosmological and local gravity
constraints \cite{Yoko,gan}.

Another representative model of dark energy based on the modification
of gravity is the Dvali-Gabadadze-Porrati (DGP) braneworld
scenario \cite{DGP}, in which the cosmic acceleration is realized
by a gravitational leakage to the extra dimension. Although this model
contains a ghost mode \cite{DGPghost} in addition to the incompatibility
with observational data \cite{DGPobser}, it has a nice feature to
recover GR in a local region through the Vainshtein mechanism \cite{Vainshtein}.
This property comes from the field self-interaction of the form $(\nabla\phi)^{2}\square\phi$,
which appears as a mixture of the transverse graviton with a brane-bending
mode \cite{DGPnon}. The non-linear field interaction suppresses the
propagation of the fifth force for the distance smaller than the so-called
Vainshtein radius $r_{V}$.

The Vainshtein mechanism was originally proposed in the context of
a Lorentz-invariant massive spin-2 Pauli-Fierz theory \cite{Pauli}.
The quadratic Pauli-Fierz theory possesses the van Dam-Veltman-Zakharov
(vDVZ) discontinuity \cite{DVZ} with which the linearized GR is not
recovered in the limit that the mass of the graviton is zero. Vainshtein
showed that in the nonlinear version of the Pauli-Fierz theory there
is a well-behaved expansion valid within a radius $r_{V}$ \cite{Vainshtein}.
Although the nonlinearities that cure the vDVZ discontinuity problem
typically give rise to the so-called Boulware-Deser ghost in massive
gravity \cite{Deser}, it is possible to construct nonlinear massive
gravitational theories free from the ghost problem in the decoupling
limit \cite{Rham}. Recently the Vainshtein mechanism was applied
to the (new) massive gravity models \cite{massiveva} and also to
Galileon models \cite{Chow,Galileonva}.

In Galileon gravity the field Lagrangian is constructed to satisfy
the Galilean symmetry $\partial_{\mu}\phi\to\partial_{\mu}\phi+b_{\mu}$
in the limit of flat spacetime \cite{Nicolis}. The nonlinear field
self-interaction $X\square\phi$, where $X=-(\nabla\phi)^{2}/2$,
appears as one of those terms \cite{DEV}. The cosmology based on
Galileon gravity has been extensively studied recently in the context
of dark energy \cite{Chow,DeFe10,Galileondark} and inflation \cite{Galileoninf}.
For the covariant Galileon there exists a de Sitter solution with
a constant field velocity. Thus the cosmic acceleration can be driven
by the field kinetic energy without a potential. Moreover there are
some viable parameter spaces in which the ghosts and Laplacian instabilities
are absent \cite{DeFe10}.

Galileon gravity can be viewed as one of the specific theories having
second-order field equations. The general action of scalar-tensor
theories with second-order equations was first derived by Horndeski
\cite{Horndeski} in the context of Lovelock gravity. This issue was
recently revisited by Deffayet \textit{et al.} \cite{Deffayet11}
as an extension of Galileon gravity (see also Ref.~\cite{Cope}).
The general $D$-dimensional action derived in Ref.~\cite{Deffayet11}
reproduces the Horndeski's action in four dimensions \cite{Koba11}.
The Galileon term $X\square\phi$, for example, can be promoted to
the form $G(\phi,X)\square\phi$, where $G$ is an arbitrary function
with respect to $\phi$ and $X$. In fact, the dynamics of dark energy
in the presence of the term $G(\phi,X)\square\phi$ have been studied
by a number of authors \cite{Galidark2}.

The Horndeski's action involves the Lagrangians of the forms ${\cal L}_{2}=P(\phi,X)$,
${\cal L}_{3}=-G(\phi,X)\square\phi$, ${\cal L}_{4}=G_{4}(\phi,X)R+G_{4,X}\times({\rm field~derivatives})$,
and ${\cal L}_{5}=G_{5}(\phi,X)G_{\mu\nu}(\nabla^{\mu}\nabla^{\nu}\phi)+(G_{5,X}/6)\times({\rm field~derivatives})$,
where $R$ is the Ricci scalar, $G_{\mu\nu}$ is the Einstein tensor,
and $G_{i,X}=\partial G_{i}/\partial X$ ($i=4,5$). For the functions
$G_{4}$ that depend on $\phi$ alone, e.g., $G_{4}=F(\phi)/2$, the
Lagrangian ${\cal L}_{4}$ reduces to ${\cal L}_{4}=F(\phi)R/2$.
For the choice $F=M_{{\rm pl}}^{2}$, where $M_{{\rm pl}}$ is the
reduced Planck mass, ${\cal L}_{4}$ corresponds to the Einstein-Hilbert
Lagrangian.

In this paper we shall study the Vainshtein mechanism in a spherically
symmetric background for the Horndeski's second-order theories with
$G_{4}=F(\phi)/2$ and $G_{5}=0$. We do not take into account the
effects of the term $G_{5}$ as well as the $X$-dependence in $G_{4}$,
but the presence of the nonlinear field interaction $G(\phi,X)\square\phi$
can allow us to understand how the Vainshtein mechanism works in the
presence of the nonminimal coupling $F(\phi)R/2$. Our analysis covers
a wide range of gravitational theories such as (extended) Galileon,
dilaton gravity, and Brans-Dicke theories with the nonlinear field
interaction.

This paper is organized as follows. In Sec.~\ref{fieldeqsec} we
derive the equations of motion for the action (\ref{action}) given
below in a spherically symmetric background. Under certain approximations
we also obtain the simplified equations for the field as well as the
gravitational potentials. In Sec.~\ref{vasec} we clarify how the
Vainshtein mechanism works in general and constrain the forms of the
action. In Sec.~\ref{appsec} the general results derived in Sec.~\ref{vasec}
are applied to specific models. We also estimate the corrections to
the gravitational potentials coming from the modification of gravity.
Sec.~\ref{consec} is devoted to conclusions.


\section{Field equations in a spherically symmetric background}

\label{fieldeqsec} 

We start with the following action 
\begin{equation}
S=\int d^{4}x\sqrt{-g}\left[\frac{1}{2}F(\phi)\, R+P(\phi,X)-G(\phi,X)\Box\phi\right]+S_{m}(g_{\mu\nu},\Psi_{m})\,,\label{action}
\end{equation}
 where $g$ is the determinant of the metric $g_{\mu\nu}$, $R$ is
the Ricci scalar, $F(\phi)$ is a function of the scalar field $\phi$,
$P(\phi,X)$ and $G(\phi,X)$ are functions of $\phi$ and $X=-g^{\mu\nu}\partial_{\mu}\phi\partial_{\nu}\phi/2$,
and $S_{m}$ is the matter action. We assume that the matter fields
$\Psi_{m}$ do not have direct couplings with the field $\phi$.

We derive the equations of motion in the spherically symmetric background
with the line element 
\begin{equation}
ds^{2}=-e^{2\Psi(r)}dt^{2}+e^{2\Phi(r)}dr^{2}+r^{2}(d\theta^{2}+\sin^{2}\theta\, d\phi^{2})\,,\label{line}
\end{equation}
 where $\Psi(r)$ and $\Phi(r)$ are functions with respect to the
distance $r$ from the center of symmetry. For the matter action $S_{m}$
we consider a perfect fluid with the energy-momentum tensor $T_{\nu}^{\mu}={\rm diag}\,(-\rho_{m},P_{m},P_{m},P_{m})$.
The $(00),(11),(22)$ components of the equations of motion derived
from the action (\ref{action}) are given, respectively, by 
\begin{eqnarray}
 &  & \left(\frac{2F}{r}+F'-2\phi'XG_{,X}\right)\Phi'+\frac{F}{r^{2}}\left(e^{2\Phi}-1\right)-F''-\frac{2F'}{r}+\phi'^{2}G_{,\phi}+2\phi''XG_{,X}=e^{2\Phi}(\rho_{m}-P)\,,\label{eq1}\\
 &  & \left(\frac{2F}{r}+F'-2\phi'XG_{,X}\right)\Psi'-\frac{F}{r^{2}}\left(e^{2\Phi}-1\right)+\frac{2F'}{r}+\phi'^{2}G_{,\phi}-\frac{4}{r}\phi'XG_{,X}=e^{2\Phi}(P_{m}+P-2XP_{,X})\,,\label{eq2}\\
 &  & F\left[\Psi''+\left(\frac{1}{r}+\frac{F'}{F}\right)(\Psi'-\Phi')+\Psi'^{2}-\Psi'\Phi'\right]+F''+\frac{F'}{r}-\phi'^{2}G_{,\phi}-2(\phi''-\Phi'\phi')XG_{,X}=e^{2\Phi}(P_{m}+P)\,,\label{eq3}
\end{eqnarray}
 where $X=-e^{-2\Phi}\phi'^{2}/2$, a prime represents the derivative
with respect to $r$, and a comma corresponds to the partial derivative
in terms of $\phi$ or $X$ (e.g., $G_{,\phi}=\partial G/\partial\phi$).
The equation of motion for the field $\phi$ is 
\begin{eqnarray}
 &  & \phi''\left[P_{,X}+2XP_{,XX}-2(G_{,\phi}+XG_{,\phi X})-2e^{-2\Phi}\phi'(G_{,X}+XG_{,XX})\left(\frac{2}{r}+\Psi'\right)\right]+e^{2\Phi}P_{,\phi}\nonumber \\
 &  & +\phi'\left[P_{,X}\left(\frac{2}{r}+\Psi'-\Phi'\right)+\phi'P_{,\phi X}-2\Phi'XP_{,XX}\right]+F_{,\phi}\left[\frac{e^{2\Phi}-1}{r^{2}}-\Psi''-\frac{2}{r}(\Psi'-\Phi')+\Psi'\Phi'-\Psi'^{2}\right]\nonumber \\
 &  & +2XG_{,X}\left(\frac{2}{r^{2}}-3\Psi'\Phi'+\Psi'^{2}+\Psi''-\frac{6}{r}\Phi'+\frac{4}{r}\Psi'\right)-4X^{2}G_{,XX}\Phi'\left(\frac{2}{r}+\Psi'\right)\nonumber \\
 &  & -2\phi'G_{,\phi}\left(\frac{2}{r}+\Psi'-\Phi'\right)-\phi'^{2}G_{,\phi\phi}+2\phi'XG_{,\phi X}\left(\frac{2}{r}+\Psi'+\Phi'\right)=0\,.\label{eq4}
\end{eqnarray}
 The continuity equation for the matter fluid is 
\begin{equation}
P_{m}'+\Psi'(\rho_{m}+P_{m})=0\,.\label{continuity}
\end{equation}
 This equation can be also derived by combining Eqs.~(\ref{eq1})-(\ref{eq4}).

In the following we focus on the weak gravitational background characterized
by the conditions $|\Phi|\ll1$ and $|\Psi|\ll1$. Then the dominant
contribution in Eq.~(\ref{eq1}) is of the order of $(F/r^{2})\Phi$.
In order to make comparisons between each term in Eqs.~(\ref{eq1})
and (\ref{eq2}) relative to $F/r^{2}$, we introduce the following
quantities 
\begin{eqnarray}
 &  & \epsilon_{F\phi}\equiv\frac{F_{,\phi}\phi'r}{F}\,,\qquad\epsilon_{GX}\equiv\frac{e^{-2\Phi}G_{,X}\phi'^{3}r}{F}\,,\qquad\epsilon_{G\phi}\equiv\frac{G_{,\phi}\phi'^{2}r^{2}}{F}\,,\qquad\epsilon_{P}\equiv\frac{e^{2\Phi}Pr^{2}}{F}\,,\nonumber \\
 &  & \epsilon_{PX}\equiv\frac{P_{,X}\phi'^{2}r^{2}}{F}\,,\qquad\epsilon_{P\phi}\equiv\frac{e^{2\Phi}P_{,\phi}\phi'r^{3}}{F}\,,\qquad\epsilon_{Pm}\equiv\frac{e^{2\Phi}P_{m}r^{2}}{F}\,,\label{epori}
\end{eqnarray}
 and 
\begin{eqnarray}
 &  & \lambda_{F\phi\phi}\equiv\frac{F_{,\phi\phi}\phi'r}{F_{,\phi}}\,,\qquad\lambda_{PX}\equiv\frac{XP_{,XX}}{P_{,X}}\,,\qquad\lambda_{PX\phi}\equiv\frac{P_{,X\phi}\phi'r}{P_{,X}}\,,\nonumber \\
 &  & \lambda_{GXX}\equiv\frac{XG_{,XX}}{G_{,X}}\,,\qquad\lambda_{G\phi X}\equiv\frac{XG_{,\phi X}}{G_{,\phi}}\,,\qquad\lambda_{G\phi\phi}\equiv\frac{G_{,\phi\phi}\phi'r}{G_{,\phi}}.
\end{eqnarray}
 The matter density $\rho_{m}$ is of the order of $(F/r^{2})\Phi$.
{}From the continuity equation (\ref{continuity}) one has 
$P_{m}/\rho_{m}\sim\Psi$
in the weak gravitational background, so that $\epsilon_{Pm}\sim\Psi^{2}$.
In the following we employ the approximation that all the terms in
Eq.~(\ref{epori}) are much smaller than 1. For the consistency with
local gravity experiments we require that these quantities are at
most of the order of $\Phi$ and $\Psi$.

{}From Eqs.~(\ref{eq1}) and (\ref{eq2}) we can express $\Phi'$
and $\Psi'$ in terms of $\rho_{m}$, $\Phi$, and (the derivatives
of) the field $\phi$. Substituting these relations into Eq.~(\ref{eq3})
and neglecting the second-order terms $\epsilon_{i}^{2}$ relative
to $\epsilon_{i}$, it follows that 
\begin{equation}
\square\Psi=\mu_{1}\rho_{m}+\mu_{2}\square\phi+\mu_{3}\,,\label{Poisson}
\end{equation}
 where $\square\equiv\dfrac{d^{2}}{dr^{2}}+\dfrac{2}{r}\dfrac{d}{dr}$,
and 
\begin{eqnarray}
\hspace{-0.7cm}\mu_{1} & \simeq & \frac{e^{2\Phi}}{4F}\left(e^{2\Phi}+1-e^{2\Phi}\epsilon_{F\phi}-e^{2\Phi}\epsilon_{GX}-\epsilon_{G\phi}+\epsilon_{P}+\epsilon_{PX}+\epsilon_{Pm}\right)\,,\label{mu1d}\\
\hspace{-0.7cm}\mu_{2} & \simeq & -\frac{(3-e^{2\Phi})(\epsilon_{F\phi}+\epsilon_{GX})}{4\phi'r}\,,\\
\hspace{-0.7cm}\mu_{3} & \simeq & -\frac{1-2e^{2\Phi}+e^{4\Phi}}{2r^{2}}\nonumber \\
 &  & -\frac{4(e^{2\Phi}-2)\epsilon_{P}-(5-3e^{2\Phi})\epsilon_{PX}+2(1-e^{2\Phi})(e^{2\Phi}\epsilon_{F\phi}+e^{2\Phi}\epsilon_{GX}+\epsilon_{G\phi})+(3-e^{2\Phi})(\lambda_{F\phi\phi}\epsilon_{F\phi}-3\epsilon_{Pm})}{4r^{2}}\,.
\end{eqnarray}
Expanding the gravitational potential $\Phi$ further and picking
up the dominant contributions, we have 
\begin{eqnarray}
\mu_{1} & \simeq & \frac{1}{4F}\left(2+6\Phi-\epsilon_{F\phi}-\epsilon_{GX}-\epsilon_{G\phi}+\epsilon_{P}+\epsilon_{PX}+\epsilon_{Pm}\right)\,,\label{mu1}\\
\mu_{2} & \simeq & -\frac{\epsilon_{F\phi}+\epsilon_{GX}}{2\phi'r}=-\frac{F_{,\phi}}{2F}+\frac{XG_{,X}}{F}\,,\label{mu2}\\
\mu_{3} & \simeq & -\frac{4\Phi^{2}+\lambda_{F\phi\phi}\epsilon_{F\phi}-2\epsilon_{P}-\epsilon_{PX}-3\epsilon_{Pm}}{2r^{2}}\,.
\end{eqnarray}
 Since $\{|\Phi|,|\epsilon_{i}|\}\ll1$, Eq.~(\ref{mu1}) gives $\mu_{1}\simeq1/(2F)$.
In GR one has $F=M_{{\rm pl}}^{2}=1/(8\pi G_{N})$ ($G_{N}$ is the
Newton's gravitational constant), so that $\mu_{1}=4\pi G_{N}$. If
$\mu_{2}\neq0$, then the field Laplacian term $\square\phi$ gives
rise to the modification to the gravitational constant. This comes
from the fact that Eq.~(\ref{eq4}) contains the matter
density $\rho_{m}$ after replacing the term $\Phi'$ by using Eq.~(\ref{eq1}).
Equation (\ref{mu2}) shows that the gravitational constant is subject
to change for the theories characterized by 
\begin{equation}
F_{,\phi}\neq0\,,\quad{\rm or}\quad G_{,X}\neq0\,.
\end{equation}
 The right hand side (r.h.s.) of Eq.~(\ref{Poisson}) is of the order of $\Phi/r^{2}$.
Provided that the condition $|\mu_{2}\phi'/r|\gg|\mu_{3}|$ is satisfied,
the contribution of the third term on the r.h.s. of Eq.~(\ref{Poisson})
can be neglected relative to the second term. This amounts to the
condition $\{|\epsilon_{F\phi}|,|\epsilon_{GX}|\}\gg\{\Phi^{2},|\lambda_{F\phi\phi}\epsilon_{F\phi}|,|\epsilon_{P}|,|\epsilon_{PX}|,|\epsilon_{Pm}|\}$.

We combine Eq.~(\ref{eq4}) with Eqs.~(\ref{eq1})-(\ref{eq3})
to derive the closed-form equation for $\phi$. Picking up the dominant
terms, we obtain 
\begin{equation}
\square\phi=\mu_{4}\,\rho_{m}+\mu_{5}\,,\label{phieq}
\end{equation}
 where 
\begin{eqnarray}
\mu_{4} & \simeq & \frac{\phi're^{2\Phi}(\alpha+\epsilon_{GX}+\epsilon_{F\phi})}{2F\alpha}\,,\label{mu4}\\
\mu_{5} & \simeq & -\phi'\{[2e^{2\Phi}(1-\lambda_{GXX})+(e^{4\Phi}-15)(1+\lambda_{GXX})]\epsilon_{GX}+2(2e^{2\Phi}-8\lambda_{G\phi X}+\lambda_{G\phi\phi}-2)\epsilon_{G\phi}\nonumber \\
 &  & -2(\lambda_{PX}e^{2\Phi}+\lambda_{PX\phi}+e^{2\Phi}-1-5\lambda_{PX})\epsilon_{PX}-2\epsilon_{P\phi}\}/(2r\alpha)\,,\label{mu5}
\end{eqnarray}
 and 
\begin{equation}
\alpha\equiv(1+\lambda_{GXX})(e^{2\Phi}+3)\epsilon_{GX}+2(1+\lambda_{G\phi X})\epsilon_{G\phi}-(1+2\lambda_{PX})\epsilon_{PX}\,.
\end{equation}
 Using the original variables with the approximation $e^{2\Phi}\simeq1$
in Eqs.~(\ref{mu4}) and (\ref{mu5}), we have 
\begin{eqnarray}
\mu_{4} & \simeq & -\frac{r(F_{,\phi}-\phi'\beta+\phi'^{2}G_{,X})}{2F\beta}\,,\label{mu4d}\\
\mu_{5} & \simeq & -\frac{P_{,\phi}r^{2}+4X(2G_{,\phi X}-P_{,XX})\phi'r+[(P_{,\phi X}-G_{,\phi\phi})r^{2}+6G_{,X}+8XG_{,XX}]\phi'^{2}}{r\beta}\,,\label{mu5d}
\end{eqnarray}
 where 
\begin{equation}
\beta\equiv(P_{,X}+2XP_{,XX}-2G_{,\phi}-2XG_{,\phi X})r-4(G_{,X}+XG_{,XX})\phi'\,.\label{betadef}
\end{equation}
 Substituting Eq.~(\ref{phieq}) into Eq.~(\ref{Poisson}), it follows
that 
\begin{equation}
\square\Psi=4\pi G_{{\rm eff}}\rho_{m}+\mu_{3}+\mu_{2}\mu_{5}\,,\label{Poimo}
\end{equation}
 where 
\begin{eqnarray}
G_{{\rm eff}} & \equiv & \frac{1}{4\pi}(\mu_{1}+\mu_{2}\mu_{4})\nonumber \\
 & \simeq & \frac{1}{8\pi F}\left[1+\left(\frac{F_{,\phi}}{2F}-\frac{XG_{,X}}{F}\right)\frac{r(F_{,\phi}-\phi'\beta+\phi'^{2}G_{,X})}{\beta}+3\Phi-\frac{1}{2}(\epsilon_{F\phi}+\epsilon_{GX}+\epsilon_{G\phi}-\epsilon_{P}-\epsilon_{PX}-\epsilon_{Pm})\right].\label{Geff}
\end{eqnarray}
 Equation (\ref{Poimo}) corresponds to the modified Poisson equation.
The second term on the r.h.s. of Eq.~(\ref{Geff}) is crucially important
for estimating the modification of gravity. For the theory in which
$P=X$ and $G=0$, the second term includes the contribution of the
order of $F_{,\phi}^{2}/(2F)$. In the presence of the dilatonic coupling
of the form $F=M_{{\rm pl}}^{2}e^{-2Q\phi/M_{{\rm pl}}}$, where $Q$
is a constant of the order of 1 and $M_{{\rm pl}}$ is the reduced
Planck mass, this contribution reduces to $2Q^{2}$ for $|\phi/M_{{\rm pl}}|\ll1$,
so that the gravitational coupling is strongly modified relative to
GR. In this case the model is in contradiction with local gravity
experiments, but the situation is different in the presence of the
term $G(\phi,X)$. 
How this modification works will be the topic of the next section.

\section{Vainshtein mechanism}
\label{vasec} 

We study how the Vainshtein mechanism works in the presence of the
non-linear field self-interaction $G(\phi,X)\square\phi$. In doing
so, we need to specify the forms of the functions $F, P, G$
at some point. Still, we aim to keep the analysis as general as possible,
so that it can cover a wide range of scalar-tensor theories.

First of all, let us consider the form of non-minimal couplings $F(\phi)$
with the Ricci scalar $R$. If we take the power-law function 
$F(\phi)\propto\phi^{p}$,
the parameter $\epsilon_{F\phi}$ reduces to $\epsilon_{F\phi}=p\phi'r/\phi$.
Suppose that there is a solution characterized by $\phi\propto r^{q}$.
Since $\epsilon_{F\phi}=pq$ in this case, the approximation $|\epsilon_{F\phi}|\ll1$
breaks down for $p$ and $q$ of the order of unity.

For example, let us consider Brans-Dicke theory \cite{Brans} characterized
by the functions $F(\phi)=\phi$ (i.e. $p=1$) and $P=\omega_{{\rm BD}}X/\phi$,
in the presence of the $X$-dependent term $G(X)$. 
For those large values of the radius characterized 
by the condition $|P_{,X}r|\gg|4(G_{,X}+XG_{,XX})\phi'|$
the term $\beta$ in Eq.~(\ref{betadef}) is given by 
$\beta\simeq\omega_{{\rm BD}}r/\phi$,
so that $\mu_{4}$ is constant for the power-law solution $\phi\propto r^{q}$.
In fact, as we will see shortly, there is a solution with $\phi\propto r^{-1}$
for constant $\mu_{4}$. In the regime $|P_{,X}r|\gg|4(G_{,X}+XG_{,XX})\phi'|$
the field equation (\ref{phieq}) is approximately given by 
\begin{equation}
\frac{d}{dr}(r^{2}\phi') \simeq \mu_{4}\rho_{m}r^{2}\,.\label{fieldeqgene}
\end{equation}
We introduce the Schwarzschild radius $r_{g}$ of the source, as
\begin{equation}
r_{g}\equiv\frac{1}{M_{{\rm pl}}^{2}}\int_{0}^{r}\rho_{m}\tilde{r}^{2}d\tilde{r}\,,\label{rg}
\end{equation}
which leads to the relation $\rho_{m}r^{2}=M_{{\rm pl}}^{2}\, dr_{g}/dr$.
Then Eq.~(\ref{fieldeqgene}) is integrated to give
\begin{equation}
\phi'(r) \simeq 
\frac{\mu_{4}M_{{\rm pl}}^{2}r_{g}}{r^{2}}\,.
\end{equation}

For the values of $r$, for which the function $r_{g}$ is almost
constant, one has $\phi'(r) \propto r^{-2}$ 
(implying $\rho_{m}r^{2}\to0$), 
which in turn leads to $\epsilon_{F\phi}=-1$. Hence the validity
of the approximation used in Sec.~\ref{fieldeqsec} breaks down for
the coupling $F(\phi)=\phi$. This problem can be avoided by rescaling
the field $\phi$ with the form of the exponential coupling, i.e.
$\phi\to M_{{\rm pl}}^{2}e^{-2Q\tilde{\phi}/M_{{\rm pl}}}$, where
$Q$ is constant. In this case the rescaled field $\tilde{\phi}$
is related with a dilaton field appearing in low-energy effective 
string theory \cite{Gas}. As we
will see in Sec.~\ref{mosec2}, after this rescaling, the kinetic
term in Brans-Dicke theory reduces to the standard one $P=-g^{\mu\nu}\partial\tilde{\phi}_{\mu}\partial\tilde{\phi}_{\nu}/2$
in the limit $Q\to0$. We treat the field $\phi$ with the exponential
coupling 
\begin{equation}
F(\phi)=M_{{\rm pl}}^{2}e^{-2Q\phi/M_{{\rm pl}}}\label{expcou}
\end{equation}
 as a more fundamental one rather than the field $\phi$ with a power-law
coupling $F(\phi)\propto\phi^{p}$. 
In this case one has $\epsilon_{F\phi}=-2Q\phi'r/M_{{\rm pl}}$,
so that $|\epsilon_{F\phi}|$ can be much smaller than 1 even for
the power-law solution like $\phi'(r) \propto r^{-2}$.

In the following we focus on the theories with the coupling (\ref{expcou}).
In doing so, we shall study two different cases: $Q\neq0$ and
$Q=0$, separately.

\subsection{$Q\neq0$}

When we discuss the case of non-zero values of $Q$, we are primarily
interested in the theories where $|Q|$ is of the order of unity. In this
case the term $F_{,\phi}$ in Eq.~(\ref{mu4d}) provides an important
contribution to the field equation. {}From Eqs.~(\ref{mu4d}) and
(\ref{mu5d}) we find that the qualitative behavior of solutions is
different depending on the radius $r$. The behavior of solutions
changes at the radius $r_{V}$ characterized by 
\begin{equation}
|B(r_{V})r_{V}|=|4(G_{,X}+XG_{,XX})(r_{V})\,\phi'(r_{V})|\,,\label{Vagene}
\end{equation}
 where 
\begin{equation}
B\equiv P_{,X}+2XP_{,XX}-2G_{,\phi}-2XG_{,\phi X}\,.
\end{equation}
 Here $r_{V}$ is the so-called Vainshtein radius \cite{Vainshtein}
below which the General Relativistic behavior can be recovered in
the presence of the term $G(\phi,X)\square\phi$.

In what follows we focus on the theories described by the action 
\begin{equation}
S=\int d^{4}x\sqrt{-g}\left[\frac{1}{2}F(\phi)R+f(\phi)X
-g(\phi)M^{1-4n} X^{n}\square\phi\right]+S_{m}(g_{\mu\nu},\Psi_{m})\,,
\label{modelcon}
\end{equation}
where $F(\phi)$ is given by Eq.~(\ref{expcou}) and $M$ is a constant
having a dimension of mass. In this case $P(\phi,X)=f(\phi)X$ and
$G(\phi,X)=g(\phi)M^{1-4n}X^{n}$, where $f(\phi),g(\phi)$ are functions
of $\phi$ and $n$ is a positive integer ($n\ge1$). We assume that
$f(\phi)$ and $g(\phi)$ are slowly varying dimensionless functions
of the order of unity, such that 
\begin{equation}
\left|M_{{\rm pl}}f_{,\phi}/f\right|\lesssim1\,,\qquad
\left|M_{{\rm pl}}g_{,\phi}/g\right|\lesssim1\,.\label{fg}
\end{equation}
If $f(\phi)$ and $g(\phi)$ are proportional to 
$F(\phi)=M_{{\rm pl}}^{2}e^{-2Q\phi/M_{{\rm pl}}}$,
the conditions (\ref{fg}) are satisfied for $|Q| \lesssim 1$.

For the action (\ref{modelcon}) Eq.~(\ref{phieq}) yields 
\begin{eqnarray}
\frac{d}{dr}(r^{2}\phi') & = & \frac{r[2QF/M_{{\rm pl}}+\{f-2(n+1)M^{1-4n}g_{,\phi}X^{n}\}\phi'r-2n(4n+1)M^{1-4n}gX^{n}]}{2F[\{f-2(n+1)M^{1-4n}g_{,\phi}X^{n}\}r-4n^{2}M^{1-4n}gX^{n-1}\phi']}\rho_{m}r^{2}\nonumber \\
 &  & -\frac{f_{,\phi}Xr^{2}+8nM^{1-4n}g_{,\phi}X^{n}\phi'r+\{(f_{,\phi}-M^{1-4n}g_{,\phi\phi}X^{n})r^{2}+2n(4n-1)M^{1-4n}gX^{n-1}\}\phi'^{2}}{\{f-2(n+1)M^{1-4n}g_{,\phi}X^{n}\}r-4n^{2}M^{1-4n}gX^{n-1}\phi'}r\,.
\label{fieldconc}
\end{eqnarray}
The qualitative behavior of the solutions to Eq.~(\ref{fieldconc})
is different depending on whether $r$ is larger than $r_{V}$ or
not. Moreover the solution is subject to change
for $r$ smaller than $r_*$, where $r_*$
is the radius at which the contribution of the density 
dependent term in Eq.~(\ref{fieldconc}) becomes 
comparable to the last term in Eq.~(\ref{fieldconc})
around a spherically symmetric body.
Hence there should
be three different regimes: (a) $r\gg r_{V}$, (b) $r_{*}\ll r\ll r_{V}$,
and (c) $r\ll r_{*}$. In the following we shall derive the solutions to
Eq.~(\ref{fieldconc}) in each regime.

\subsubsection{$r\gg r_{V}$}

As long as the condition $|F_{,\phi}|\gg|\phi'\beta|$ is satisfied
in the regime $r\gg r_{V}$, the term $\mu_{4}$ in Eq.~(\ref{mu4d})
can be estimated as $\mu_{4}\simeq Q/(M_{{\rm pl}}B)$. As we will
see below, the field behaves as $\phi'(r)\simeq(QM_{{\rm pl}}/B)(r_{g}/r^{2})$
in the regime $r\gg r_{V}$ for $B\simeq{\rm constant}$. In this
case the condition $|F_{,\phi}|\gg|\phi'\beta|$ is in fact satisfied.
For the theories given by (\ref{modelcon}) the term $\mu_{4}\rho_{m}$
is the dominant contribution to the r.h.s. of Eq.~(\ref{phieq})
for $r\gg r_{V}$. This is known by deriving the solution of Eq.~(\ref{phieq})
without the term $\mu_{5}$ and by substituting the solution into
Eqs.~(\ref{mu4d}) and (\ref{mu5d}). Then, in the regime $r\gg r_{V}$,
the field equation (\ref{fieldconc}) reduces to 
\begin{equation}
\frac{d}{dr}(r^{2}\phi')\simeq\frac{Q}{M_{{\rm pl}}B}\rho_{m}r^{2}\,,
\label{fieldgene}
\end{equation}
where $B=f-2(n+1)M^{1-4n}g_{,\phi}X^{n}\simeq f$. 
As long as $f(\phi)$ is a slowly varying function,
$B$ is nearly constant. In general, for k-essence theories
in which $P$ includes non-linear terms in $X$ \cite{kes}, 
$B$ depends on $r$.
Provided that the term of the form $P=f(\phi)X$ corresponds to the
dominant contribution to $B$, one can also employ the approximation
that $B$ is nearly constant. Using the Schwarzschild radius defined
in Eq.~(\ref{rg}), Eq.~(\ref{fieldgene}) is integrated to give
\begin{equation}
\phi'(r)\simeq\frac{QM_{{\rm pl}}}{B}\frac{r_{g}}{r^{2}}\,.\label{phipgene}
\end{equation}
Here we neglected the solution of the homogeneous
differential equation
as this is equivalent
to the renormalization of $r_{g}$. Under the approximation that the
solution (\ref{phipgene}) is valid at $r=r_{V}$, it follows that
\begin{equation}
r_{V}^{3}\simeq\left|\frac{4QM_{{\rm pl}}r_{g}}{B(r_{V})^{2}}(G_{,X}+XG_{,XX})(r_{V})\right|\,.\label{rvdef}
\end{equation}

For the function $G=X/M^{3}$ the Vainshtein radius is known to be
$r_{V}=|4QM_{{\rm pl}}r_{g}/(B(r_{V})^{2}M^{3})|^{1/3}$. If the term
$G_{,X}+XG_{,XX}$ depends on $X$, we need to use Eq.~(\ref{phipgene})
again to derive the closed-form expression of $r_{V}$.

\subsubsection{$r_{*}\ll r\ll r_{V}$}

In the regime $r_{*}\ll r\ll r_{V}$ the $G$-dependent terms are
the dominant contributions to $\beta$. As long as the solution is
described by $\phi'(r) \propto r^{-p}$ with $0<p<1$, 
one can approximate $\beta\simeq-4(G_{,X}+XG_{,XX})\phi'(r)$. 
For the function $G=g(\phi)M^{1-4n}X^{n}$ ($n \geq 1$) 
the solution derived later behaves as $\phi'(r)\propto r^{-1/(2n)}$,
so that the approximation given above is justified. In the regime
$r_{*}\ll r\ll r_{V}$ the terms $(6G_{,X}+8XG_{,XX})\phi'^{2}$ in
the numerator of $\mu_{5}$ in Eq.~(\ref{mu5d}) provide the dominant
contribution to the field equation, and hence 
\begin{equation}
\frac{d}{dr}(r^{2}\phi')\simeq\frac{3G_{,X}+4XG_{,XX}}
{2(G_{,X}+XG_{,XX})}\, r\phi'=\frac{4n-1}{2n}\, r\phi'\,.\label{phieqgene}
\end{equation}
Since $\lambda_{GXX}=XG_{,XX}/G_{,X}=n-1$, the coefficient in front
of the term $r\phi'$ in Eq.~(\ref{phieqgene}) is constant for the
function $G=g(\phi)M^{1-4n} X^{n}$. 
Equation (\ref{phieqgene}) is integrated to give 
\begin{equation}
\phi'(r)=Cr^{-1/(2n)}\,.\label{phipgene2}
\end{equation}
The coefficient $C$ is approximately known by matching two solutions
(\ref{phipgene}) and (\ref{phipgene2}) at $r=r_{V}$, which gives
$C=QM_{{\rm pl}}r_{g}r_{V}^{1/(2n)-2}/B(r_{V})$. Then Eq.~(\ref{phipgene2})
reduces to 
\begin{equation}
\phi'(r)\simeq\frac{QM_{{\rm pl}}r_{g}}{B(r_{V})r_{V}^{2}}
\left(\frac{r}{r_{V}}\right)^{-1/(2n)}\,.\label{phiva}
\end{equation}
Compared to the solution (\ref{phipgene}) the field derivative varies
more slowly in the regime $r_{*}\ll r\ll r_{V}$. 
This is the region in which the Vainshtein mechanism is at work.
The solution (\ref{phiva}) is compatible with the approximations 
we made to find Eq.~(\ref{phieqgene}), e.g.,
$|g_{,\phi}g^{-1}\phi'r|\ll1$. 

The solution (\ref{phiva}) diverges in the limit $r\to0$.
To avoid this divergent behavior we expect that $\phi' (r)$ behaves in
a different way for the radius smaller than $r_{*}$. In order to find
the radius $r_{\epsilon}$ below which the approximation 
we used breaks down, we compute the variables 
defined in Eq.~(\ref{epori}):
\begin{eqnarray}
\hspace{-0.8cm} &  & \epsilon_{F\phi}\simeq-\frac{2Q^{2}}{B(r_{V})}\frac{r_{g}}{r_{V}}\left(\frac{r}{r_{V}}\right)^{1-1/(2n)},\qquad|\epsilon_{GX}|\simeq\frac{M_{{\rm pl}}^{2}}{8nF(\phi)}\frac{r_{g}}{r}|\epsilon_{F\phi}|,\qquad|\epsilon_{G\phi}|\simeq\frac{|Q^{3}\eta_{g}(\phi)M_{{\rm pl}}^{2}|}{8n^{2}B(r_{V})^{2}F(\phi)}\left(\frac{r_{g}}{r_{V}}\right)^{3}\left(\frac{r}{r_{V}}\right)^{1-1/n},\nonumber \\
\hspace{-0.8cm} &  & \epsilon_{PX}=-2\epsilon_{P}\simeq\frac{f(\phi)Q^{2}M_{{\rm pl}}^{2}}{B(r_{V})^{2}F(\phi)}\left(\frac{r_{g}}{r_{V}}\right)^{2}\left(\frac{r}{r_{V}}\right)^{2-1/n}\,,\qquad\epsilon_{P\phi}\simeq\frac{f(\phi)\eta_{f}(\phi)Q^{3}M_{{\rm pl}}^{2}}{2B(r_{V})^{3}F(\phi)}\left(\frac{r_{g}}{r_{V}}\right)^{3}\left(\frac{r}{r_{V}}\right)^{3-3/(2n)},\label{epex}
\end{eqnarray}
 where $\eta_{f}(\phi)=M_{{\rm pl}}f_{,\phi}/f$ and 
 $\eta_{g}(\phi)=M_{{\rm pl}}g_{,\phi}/g$.
As long as $F(\phi)/M_{\rm pl}^2$, $f(\phi)$, $g(\phi)$ do not
change significantly and they remain of the order of unity, the quantities
given in Eq.~(\ref{epex}), apart from $|\epsilon_{GX}|$, are much
smaller than 1 for $r\ll r_{V}$. The variable $\epsilon_{GX}$ is
proportional to $r^{-1/(2n)}$, which diverges in the limit $r\to0$.
The validity of the approximation $|\epsilon_{GX}|\ll1$ breaks
down for the radius $r<r_{\epsilon}$, where 
\begin{equation}
r_{\epsilon}=r_{V}\left(\frac{M_{{\rm pl}}^{2}Q^{2}}{4nF(r_{\epsilon})|B(r_{V})|}
\frac{r_{g}^{2}}{r_{V}^{2}}\right)^{2n}\,.\label{rep}
\end{equation}
For $|Q|$, $|B(r_{V})|$, and $F(r_{\epsilon})/M_{{\rm pl}}^{2}$
of the order of unity one has 
$r_{\epsilon}/r_{g} \approx (r_{g}/r_{V})^{4n-1}$.
This shows that, for $n\geq1$ and $r_{g}\ll r_{V}$, 
$r_{\epsilon}$ is extremely small even compared to $r_{g}$.
As we will see in Sec.~\ref{appsec} the typical Vainshtein 
radius for the Sun ($r_g \approx 10^5$\,cm) is around 
$r_V \approx 10^{20}$\,cm for the models relevant to 
dark energy.
When $n=1$ one has $r_{\epsilon} \approx 10^{-40}$\,cm, 
which is even smaller than the Planck length.
As we will see in Sec.~\ref{appsec} the typical value of 
$r_*$ is about the radius of the Sun ($\approx 10^{10}$ cm), 
so that the solution (41) is trustable for $r>r_*$.

\subsubsection{$r\ll r_{*}$}

Let us derive the solution to the field equation in the regime 
where the density dependent term on the r.h.s. of Eq.~(\ref{fieldconc})
becomes important around the spherical symmetric body.
For the regularity of solutions the boundary conditions should satisfy 
$\phi'(0)=0$ and $|\phi''(0)|<\infty$ at the origin.
These two conditions lead to $\phi'(r)\propto r^{m}$ ($m\geq1$),
as $r\to0$. We also impose that the density $\rho_{m}$ approaches
a constant value $\rho_{c}$ in the limit $r\to0$. 

For the theories with $n=1$, i.e. $G=g(\phi) M^{-3}$, Eq.~(\ref{fieldconc})
around $r=0$ reads 
\begin{equation}
\frac{d}{dr}(r^{2}\phi')\simeq\frac{M^{3}Q\rho_{m}}{M_{{\rm pl}}
(rM^{3}f_{c}-4g_{c}\phi')}r^{3}-\frac{6\phi'^{2}g_{c}}
{rM^{3}f_{c}-4g_{c}\phi'}r\,,\label{phisma}
\end{equation}
where $f_{c}=f(\phi_{c})$ and $g_{c}=g(\phi_{c})$ with $\phi_{c}$
being the field value at the origin. 
On using $\rho_{m}\simeq\rho_{c}={\rm constant}$,
there is a solution characterized by $\phi'(r)=br$. The coefficient
$b$ is known by substituting $\phi'(r)=br$ into Eq.~(\ref{phisma}).
This leads to the following solution 
\begin{equation}
\phi'(r)\simeq\frac{M^{3}f_{c}}{4g_{c}}\left[1\pm
\sqrt{1-\frac{8Q\rho_{c}g_{c}}{3M_{{\rm pl}}M^{3}f_{c}^{2}}}\right]r\,.
\label{phili}
\end{equation}
If the condition $|Q\rho_{c}g_{c}|\gg M_{{\rm pl}}M^{3}f_{c}^{2}$
is satisfied, Eq.~(\ref{phili}) is approximately given by 
\begin{equation}
\phi'(r)\simeq\pm\left(\frac{|Q|M^{3}\rho_{c}}{6M_{{\rm pl}}g_{c}}
\right)^{1/2}r\,,
\label{phili2}
\end{equation}
whose existence requires that $Q<0$ for $g_{c}>0$. The sign of
Eq.~(\ref{phili2}) is fixed by matching this solution with the one
in the regime $r\gg r_{*}$. When $B(r_{V})>0$ and $B(r_{V})<0$
the sign of Eq.~(\ref{phili2}) is negative and positive, respectively, 
for $Q<0$. 

While we derived the solution (\ref{phili2}) around the center 
of the star, this is also valid for the star where 
$\rho_m$ is approximately constant.
Matching the two solutions (\ref{phiva}) and (\ref{phili2}) 
at the radius $r_*$ for $n=1$, it follows that 
\begin{equation}
r_* \simeq \left( \frac{6|Q g_c|}{B^2(r_V)}
\frac{r_g^2}{\rho_c r_V^3} \right)^{1/3}
\frac{M_{\rm pl}}{M}\,,
\label{rstar}
\end{equation}
at which the two terms on the r.h.s of Eq.~(\ref{fieldconc})
are the same order.
The matching radius $r_*$ depends on $M$
as well as $\rho_c$.
For the models relevant to the cosmic acceleration today,
$r_*$ for the Sun is typically around its radius
(provided that the density of the Sun is assumed 
to be nearly constant).

For $n>1$ the field equation satisfying the
boundary conditions at $r=0$ reduces to 
\begin{equation}
\frac{d}{dr}(r^{2}\phi')\simeq\frac{Q\rho_{c}}{M_{{\rm pl}}f_{c}}r^{2}\,.
\end{equation}
This is integrated to give 
\begin{equation}
\phi'(r)\simeq\frac{Q\rho_{c}}{3M_{{\rm pl}}f_{c}}r\,,
\label{phili3}
\end{equation}
where we used $\phi'(0)=0$.
In order to match this solution with (\ref{phiva}) we require that 
$B(r_{V})>0$.

\subsubsection{Corrections to the gravitational potentials}

We estimate the modifications to the Newtonian gravitational potentials
in the regime $r_{*}\ll r\ll r_{V}$. First of all, let us see how
the Vainshtein mechanism suppresses the additional gravitational coupling
appearing in Eq.~(\ref{Geff}). Since $|\epsilon_{GX}|\ll|\epsilon_{F\phi}|$
for $r_{*}\ll r\ll r_{V}$, the term $XG_{,X}/F$ can be neglected relative
to $F_{,\phi}/(2F)$. Using the solution (\ref{phiva}) and the definition
of $r_{V}$ given in Eq.~(\ref{rvdef}), it follows that 
$|\beta|\simeq|4(G_{,X}+XG_{,XX})\phi'|\simeq|B(r_{V})|r_{V}(r_{V}/r)^{1-1/(2n)}$.
Then the second term in the square bracket of Eq.~(\ref{Geff}) can
be estimated as 
\begin{equation}
\left|\left(\frac{F_{,\phi}}{2F}-\frac{XG_{,X}}{F}\right)\frac{r(F_{,\phi}-\phi'\beta+\phi'^{2}G_{,X})}{\beta}\right|\simeq\left|\frac{rF_{,\phi}^{2}}{2F\beta}\right|\simeq\frac{2Q^{2}}{M_{\rm pl}^2}\left|\frac{F(\phi)}{B(r_{V})}\right|\left(\frac{r}{r_{V}}\right)^{2-1/(2n)}\,,
\end{equation}
which is much smaller than 1 for $r\ll r_{V}$. Hence the presence
of the term $G(\phi,X)$ in $\beta$ can lead to the recovery of GR
within the Vainshtein radius.

Using the estimation (\ref{epex}) and picking up the dominant contributions
in Eqs.~(\ref{eq1}) and (\ref{eq2}), it follows that 
\begin{eqnarray}
 &  & \frac{2F}{r}\Phi'+\frac{2F}{r^{2}}\Phi-F''-\frac{2F'}{r}\simeq\rho_{m}\,,
 \label{graeq1}\\
 &  & \frac{2F}{r}\Psi'-\frac{2F}{r^{2}}\Phi+\frac{2F'}{r}\simeq0\,.\label{graeq2}
\end{eqnarray}
Substituting the solution (\ref{phiva}) into Eq.~(\ref{graeq1}),
we obtain 
\begin{equation}
\frac{d}{dr}\left(r\Phi-\frac{r_{g}}{2}\right)\simeq
\frac{(1-4n)Q^{2}r_{g}}{2nB(r_{V})r_{V}^{2-1/(2n)}}r^{1-1/(2n)}\,.
\end{equation}
Integration of this equation leads to 
\begin{equation}
\Phi\simeq\frac{r_{g}}{2r}\left[1-\frac{2Q^{2}}{B(r_{V})}\left(\frac{r}{r_{V}}\right)^{2-1/(2n)}\right]\,,\label{Phiso}
\end{equation}
where we neglected the homogeneous solution, as it corresponds
to the renormalization of $r_{g}$. 
Plugging this solution into Eq.~(\ref{graeq2}), we get 
\begin{equation}
\Psi\simeq-\frac{r_{g}}{2r}\left[1-\frac{4n}{2n-1}\frac{Q^{2}}{B(r_{V})}\left(\frac{r}{r_{V}}\right)^{2-1/(2n)}\right]\,.\label{Psiso}
\end{equation}
 Clearly the second terms on the r.h.s. of the square brackets of 
 Eqs.~(\ref{Phiso}) and (\ref{Psiso}) are much smaller than unity 
 in the regime $r\ll r_{V}$,
so that the fifth force is suppressed.

We define the post-Newtonian parameter $\gamma$, as 
\begin{equation}
\gamma\equiv-\Phi/\Psi\,.
\end{equation}
 The present tightest experimental bound on $\gamma$ is $|\gamma-1|<2.3\times10^{-5}$
\cite{Will}. Using the solutions (\ref{Phiso}) and (\ref{Psiso})
this constraint translates into 
\begin{equation}
\frac{2Q^{2}}{2n-1}\frac{1}{|B(r_{V})|}\left(\frac{r}{r_{V}}\right)^{2-1/(2n)}<2.3\times10^{-5}\,.\label{bound}
\end{equation}
For $r$ much less than $r_{V}$ the bound (\ref{bound}) 
can be satisfied even for $|Q|={\cal O}(1)$.

In the regime $r\ll r_{*}$ the corrections to the gravitational potentials
are even more suppressed than those estimated by Eqs.~(\ref{Phiso})
and (\ref{Psiso}). This comes from the fact that, as $r$ approaches 0, 
the quantities such as $|\epsilon_{F\phi}|$ and $|\epsilon_{GX}|$ decrease.

\subsection{$Q=0$}

\label{Qzero} Let us study the theories given by the action (\ref{modelcon})
with $F=M_{{\rm pl}}^{2}$. Even for $Q=0$ there is
an additional correction term $XG_{,X}/F$ in Eq.~(\ref{Geff}),
so it is not clear whether such a correction is suppressed or not.
We discuss two different cases: (i) $n=1$ and (ii) $n>1$.

\subsubsection{$n=1$}

This case corresponds to the function $G=g(\phi)M^{-3}X$. The qualitative
behavior of solutions changes at the radius $r_{V}$ characterized
by the condition $|f_{V}r_{V}|=|4M^{-3}g_{V}\phi'(r_{V})|$, 
where $f_{V}$ and $g_{V}$
are the values of $f$ and $g$ at $r=r_{V}$, respectively.

For $r\gg r_{V}$ the term $\mu_{4}\rho_{m}$ dominates over $\mu_{5}$
in Eq.~(\ref{phieq}), where $\mu_{4}\simeq\phi'r/(2M_{{\rm pl}}^{2})$.
It then follows that 
\begin{equation}
\frac{d}{dr}(r^{2}\phi')=\frac{\phi'r}{2}\frac{dr_{g}}{dr}\,.
\end{equation}
The solution to this equation can be written in the form 
\begin{equation}
\phi'(r)=\frac{C}{r^{2}}\exp\left[\frac{1}{2M_{{\rm pl}}^{2}}\int_{r_{V}}^{r}
\rho_{m}(\tilde{r})\tilde{r}d\tilde{r}\right]\,,\label{phin1}
\end{equation}
where $C$ is an integration constant. For the local matter density
we assume that the integral 
$\int^{\infty}\rho_{m}(\tilde{r})\,\tilde{r}^{2}d\tilde{r}$
is finite. Then, for large $r$, we require $\rho_{m}\simeq br^{-2-q}$,
with $q>1$ ($b$ is a constant). In this case Eq.~(\ref{phin1})
yields 
\begin{equation}
\phi'(r)=\frac{C}{r^{2}}\exp\left[-\frac{b}{2M_{{\rm pl}}^{2}q}(r^{-q}-r_{V}^{-q})\right]\simeq\frac{C_{1}}{r^{2}}\,,\label{eq:oVa}
\end{equation}
where $C_{1}$ is another constant which absorbs the exponential
term (which is nearly constant for large $r$). Note that this relation
is valid even for the weaker bound $q>0$.

Let us consider the regime $r<r_V$. Around the
center of the spherical symmetry there should be the change of solutions
at some radius $r_{*}$, so we first derive the solution in the regime
$r_{*} \ll r\ll r_{V}$. Since $\beta\simeq-4M^{-3}g\phi'$, 
$\mu_{4}\simeq5\phi'r/(8M_{{\rm pl}}^{2})$,
and $\mu_{5}\simeq3\phi'/(2r)$ in this region, the field equation is 
\begin{equation}
\frac{d}{dr}(r^{2}\phi')=\phi'r\left(\frac{5}{8}\frac{dr_{g}}{dr}+\frac{3}{2}\right)\,.
\end{equation}
Under the condition $|dr_{g}/dr|\ll1$ (which corresponds to $\rho_{m}r^{2}\to0$
for large $r$) we have the approximate solution 
\begin{equation}
\phi'(r)\simeq\frac{C_{2}}{\sqrt{r}}\,,\label{eq:iVa}
\end{equation}
where $C_{2}$ is an integration constant. 
The solution (\ref{eq:iVa})
cannot be trusted up to $r\to0$.

Finally we study the behavior of the solution in the region $r \ll r_{*}$. 
For the regularity at the origin the field derivative should take the
form $\phi'(r)\propto r^{m}$, where $m\geq1$. 
Since $\rho_{m}=\rho_{c}+{\cal O}(r^{2})$ in this regime,
the field equation (\ref{fieldconc}) is approximately given by 
\begin{equation}
\frac{d}{dr}(r^{2}\phi')\simeq\frac{6M^{-3}g_{c}{\phi'}^{2}r}
{4M^{-3}g_{c}\phi'-f_{c}r}\,.\label{fiorigin}
\end{equation}
Assuming the solution of the form $\phi'=b\, r$
and substituting it into Eq.~(\ref{fiorigin}), we find the following
solution 
\begin{equation}
\phi'(r)\simeq\frac{f_{c}}{2g_{c}}M^{3}r\,.\label{fiorigin2}
\end{equation}

Let us match Eq.~(\ref{fiorigin2})
with Eq.~(\ref{eq:iVa}) at the radius $r=r_{*}$. 
We caution that, if the density $\rho_m$ differs from $\rho_c$
at $r=r_{*}$, the correction to the solution (\ref{fiorigin2})
should be taken into account. 
Since $C_{2}=f_{c}/(2g_{c})\, M^{3}\, r_{*}^{3/2}$ after 
the matching, the solution in the regime $r_{g}<r\ll r_{V}$ is 
\begin{equation}
\phi'(r)\simeq\frac{f_{c}}{2g_{c}}M^{3}r_{*}
\left(\frac{r_{*}}{r}\right)^{1/2}\,.\label{eq:iVa2}
\end{equation}
Finally we match Eq.~(\ref{eq:iVa2}) with Eq.~(\ref{eq:oVa})
at $r\simeq r_{V}$. This leads to the following solution in the regime
$r\gg r_{V}$: 
\begin{equation}
\phi'(r)\simeq\frac{f_{c}}{2g_{c}}M^{3}r_{*}
\left(\frac{r_{*}}{r_{V}}\right)^{1/2}\left(\frac{r_{V}}{r}\right)^{2}\,.
\label{eq:iVa3}
\end{equation}
The Vainshtein radius $r_V$ is defined by the condition 
$|f_{V}r_{V}|\simeq|4M^{-3}g_{V}\phi'(r_{V})|$. 
This gives
\begin{equation}
r_{V}\simeq\left|\frac{2f_{c}\,g_{V}}{f_{V}g_{c}}\right|^{2/3}r_{*}\,.\label{rvrs}
\end{equation}
For $|f|$ and $|g|$ of the order of unity, Eq.~(\ref{rvrs}) implies
that $r_{V}$ is the same order as $r_{*}$. This means that there
is no intermediate regime $r_{*}<r<r_{V}$ characterized by the solution
(\ref{eq:iVa2}). Moreover the matching radius $r_{V}$ cannot be
fixed completely.

Let us consider the regime $r<r_{V}$ with the solution
(\ref{fiorigin2}). Then the second term in the square bracket of
Eq.~(\ref{Geff}) can be estimated as 
\begin{equation}
\xi\equiv\left|\left(\frac{F_{,\phi}}{2F}-\frac{XG_{,X}}{F}\right)\frac{r(F_{,\phi}-\phi'\beta+\phi'^{2}G_{,X})}{\beta}\right|\approx|\epsilon_{GX}|\approx\frac{M^{6}r^{4}}{M_{{\rm pl}}^{2}}\,.
\end{equation}
If the same model is responsible for the late-time cosmic acceleration,
the mass $M$ is related with today's Hubble parameter $H_{0}$ as
$M^{3}\approx M_{{\rm pl}}H_{0}^{2}$. Using this relation one has
$\xi\approx(r/H_{0}^{-1})^{4}$, which means that the correction is 
significantly suppressed on solar-system scales.
The $G$-dependent terms appearing in Eqs.~(\ref{eq1})
and (\ref{eq2}) are the orders 
of $(F/r^{2})\epsilon_{GX}$ and $(F/r^{2})\epsilon_{G\phi}$.
For the radius $r<r_{V}$ we have $|\epsilon_{G\phi}|\approx(r/H_0^{-1})^{6}$,
which is even much smaller than $|\epsilon_{GX}|$. Then the corrections
to the gravitational potentials coming from the term $G(\phi,X)\square\phi$
are strongly suppressed on solar-system scales.

In the regime $r>r_V$ the field derivative $\phi'(r)$ is a decreasing 
function with respect to $r$, so the correction $\xi$ becomes
maximum around $r=r_V$. As long as $r_V \ll H_0^{-1}$, 
$\xi(r_V) \approx (r_V/H_0^{-1})^4$ is much smaller than 1.
Of course, if $r_* (\sim r_V)$ is significantly away from the origin,
we need to take into account the correction to the solution 
(\ref{fiorigin2}) coming from the change of the matter density.
Still the small cubic mass term $M^3$ appearing in Eq.~(\ref{fiorigin2})
would affect the solutions in the regime $r>r_V$ [as it happens in 
Eqs.~(\ref{eq:iVa2}) and (\ref{eq:iVa3})], 
so that the term $\xi$ should be suppressed as well.

\subsubsection{$n>1$}

We proceed to the case in which $n>1$. Let us study the behavior
of solutions around the origin. For the regularity we need to assume
the form $\phi' (r) \propto r^{m}$ ($m\geq1$) and $\rho_{m}\to\rho_{c}$,
as $r\to0$. Then Eq.~(\ref{fieldconc}) yields 
\begin{equation}
\frac{d}{dr}(r^{2}\phi')=\frac{\rho_{c}}{2M_{{\rm pl}}^{2}}\, r^{3}\phi'
-\frac{f_{,\phi}(\phi_{c})}{2f(\phi_{c})}\, r^{2}\phi'^{2}\,.\label{fieldconc2}
\end{equation}
For the theories in which $f$ is constant the solution to Eq.~(\ref{fieldconc2})
is given by $\phi'(r)\propto e^{\rho_{c}r^{2}/(4M_{{\rm pl}}^{2})}/r^{2}$,
which is singular at $r=0$. If $f$ depends on $\phi$, we obtain
the following solution 
\begin{equation}
\phi'(r)\simeq-\frac{2f(\phi_{c})}{f_{,\phi}(\phi_{c})}\frac{1}{r}\,,
\end{equation}
which is again singular at $r=0$. 
In both cases the solutions cannot satisfy the regularity condition 
at the origin. The theories
with $n>1$ and $Q=0$ are not viable because of the above mentioned
property.

\section{Application to concrete models}
\label{appsec} 

In this section we apply our formulas given
in Sec.~\ref{vasec} to a number of concrete models.

\subsection{Extended Galileon}
\label{mosec1} 

We first study the theories characterized by 
\begin{equation}
F(\phi)=M_{{\rm pl}}^{2}e^{-2Q\phi/M_{{\rm pl}}}\,,\qquad 
P(X)=\epsilon X\,,\qquad G(X)=\lambda M^{1-4n}X^{n}\,,
\label{egalileon}
\end{equation}
where $\epsilon=\pm 1$, $n$ is a positive integer ($n\geq1$), and 
$\lambda$ is a constant of the order of unity (which can be either 
positive or negative).
In this case the functions $f(\phi)$ and $g(\phi)$ in Eq.~(\ref{modelcon})
are strictly constant, i.e. $f(\phi)=\epsilon$ and $g(\phi)=\lambda$.

The covariant Galileon model, which recovers the Galilean symmetry
$\partial_{\mu}\phi\to\partial_{\mu}\phi+b_{\mu}$ in the limit of
Minkowski spacetime, corresponds to $n=1$ \cite{Nicolis,DEV}. 
Note that in the DGP model the field self-interaction
of the form $\lambda M^{-3}X\square\phi$ arises from a brane-bending mode.
For general $n$ the background expansion of the Universe is the same
as that of the Dvali-Turner model \cite{Kimura}. 
When $\epsilon=-1$, $Q=0$, and $\lambda>0$ 
there is a de Sitter attractor along which $\dot{\phi}={\rm constant}$.
If this solution is responsible for the cosmic acceleration today,
the mass $M$ is related to the today's Hubble radius 
$r_{c}=H_{0}^{-1} \approx 10^{28}\,$cm
via $M\approx(M_{{\rm pl}}^{1-2n}r_{c}^{2n})^{1/(1-4n)}$ 
\cite{Kimura,Defe2011}.

Let us consider the case $Q\neq0$. Using Eq.~(\ref{phipgene}),
the Vainshtein radius $r_{V}$ defined in Eq.~(\ref{rvdef}) reads
\begin{equation}
r_{V}=(2^{3-n}n^{2} |\lambda| )^{1/(4n-1)}
\frac{(|Q|M_{{\rm pl}}r_{g}
)^{(2n-1)/(4n-1)}}{M}
\approx\frac{(|Q|M_{{\rm pl}}r_{g})^{(2n-1)/(4n-1)}}{M}\,.
\end{equation}
If the mass $M$ has an approximate relation 
$M\approx(M_{{\rm pl}}^{1-2n}r_{c}^{2n})^{1/(1-4n)}$
(as in the case of $Q=0$), one has 
$r_{V}\approx(|Q|r_{g}^{2n-1}r_{c}^{2n})^{1/(4n-1)}$.
When $n=1$ this reduces to $r_{V}\approx(|Q|r_{g}r_{c}^{2})^{1/3}$,
which recovers the Vainshtein radius $r_{V}\approx(r_{g}r_{c}^{2})^{1/3}$
in the DGP model for $|Q|={\cal O}(1)$.
For the Sun ($r_{g} \approx 10^{5}$\,cm) 
one has $r_{V}\approx10^{20}$\,cm
for $|Q|={\cal O}(1)$.

From Eqs.~(\ref{phipgene}) and (\ref{phiva}) the solutions in the
regimes $r\gg r_{V}$ and $r_{*}\ll r\ll r_{V}$ are given, respectively, 
by $\phi'(r)\simeq QM_{{\rm pl}}r_{g}/(\epsilon r^{2})$
and $\phi'(r)\simeq QM_{{\rm pl}}r_{g}/(\epsilon r_{V}^{2})
(r/r_{V})^{-1/(2n)}$.
If we consider the case $n=1$ and $\lambda>0$
with the condition $|Q\rho_{c} \lambda|\gg M_{{\rm pl}}M^{3}$,
the solution in the regime $r\ll r_{*}$
is $\phi'(r)\simeq-[|Q|M^{3}\rho_{c}/(6M_{{\rm pl}} 
\lambda)]^{1/2}r$ for $\epsilon=+1$ and 
$\phi'(r)\simeq[|Q|M^{3}\rho_{c}/
(6M_{{\rm pl}} \lambda)]^{1/2}r$
for $\epsilon=-1$ respectively 
(where in both cases $Q<0$).

For $n=1$, the matching radius $r_{*}$ given in 
Eq.~(\ref{rstar}) can be estimated as
\begin{equation}
\frac{r_{*}}{r_{g}}\simeq\left(\frac{6|Q \lambda|}
{\rho_{c}r_{V}^{4}}\frac{r_{V}}{r_{g}}
\frac{M_{{\rm pl}}^{3}}{M^{3}}\right)^{1/3}\,.\label{rr}
\end{equation}
If the mass $M$ is related to $r_{c}$ via $M^{3}\approx M_{{\rm pl}}r_{c}^{-2}$
then the ratio (\ref{rr}) yields $r_{*}/r_{g}\approx[2|Q \lambda| (r_{V}/r_{g})
(\rho_{0}/\rho_{c})
(r_{c}/r_{V})^{4}]^{1/3}$, where $\rho_{0}\approx3M_{{\rm pl}}^{2}/r_{c}^{2}
\approx10^{-29}$\,g/cm$^{3}$ is the cosmological density today. 
For the Sun ($\rho_{c}\approx10^{2}$\,g/cm$^{3}$) one has
$r_{*}\approx10^{5}r_{g}\approx10^{10}$\,cm for $|Q \lambda|={\cal O}(1)$,
which is the same order as the radius of the Sun. 
The distance $r_{\epsilon}$ given in Eq.~(\ref{rep})
is very much smaller than $r_{*}$, so that the solutions derived
in Sec.~\ref{vasec} are trustable.
We caution that around $r=r_*$ there is a correction to 
the solution (\ref{phili2}) coming from the varying matter density, 
but still $r_*$ cannot be smaller than $r_{\epsilon}$.

When $n>1$, matching two solutions in the regimes $r\ll r_{*}$ and
$r_{*}\ll r\ll r_{V}$ requires that $B(r_{V})>0$ and hence $\epsilon>0$.
However the existence of a late-time de Sitter solution 
($\dot{\phi}={\rm constant}$) requires $\epsilon<0$, 
which means that there are no solutions with appropriate 
boundary conditions around the origin.
While the model with positive $\epsilon$ may be irrelevant to dark energy, 
it does not possess
the discontinuous behavior. 

If $n=1$, the gravitational potentials (\ref{Phiso}) and (\ref{Psiso})
in the regime $r_{*}\ll r\ll r_{g}$ are given by 
\begin{equation}
\Phi\simeq\frac{r_{g}}{2r}\left[1-\frac{2Q^{2}}{\epsilon}
\left(\frac{r}{r_{V}}\right)^{3/2}\right]\,,
\qquad\Psi\simeq-\frac{r_{g}}{2r}\left[1-\frac{4Q^{2}}{\epsilon}
\left(\frac{r}{r_{V}}\right)^{3/2}\right]\,.\label{phiexa}
\end{equation}
When $|Q| \lesssim 1$ the experimental bound (\ref{bound}) on
the post-Newtonian parameter $\gamma$ is satisfied for 
$r<5\times10^{-4}\, r_{V}$.
If the relation $M^{3}\approx M_{{\rm pl}}r_{c}^{-2}$ holds, this
bounds translates into $r<10^{17}$\,cm for the Sun and $r<10^{15}$\,cm
for the Earth. Note that for the radius $r\ll r_{*}$ the corrections
to $\Phi$ and $\Psi$ are much smaller than those given in Eq.~(\ref{phiexa}).
Hence the model is compatible with the experimental bound 
on the solar-system scales.

When $Q=0$ we showed in Sec.~\ref{Qzero} that the corrections to
the gravitational potentials are extremely tiny for $n=1$, so that
the model can pass the solar-system constraints. The models with $n>1$
are plagued by the problem of the singularity of $\phi'(r)$ at the
origin. We note that this situation may change in the presence of
other non-linear field corrections appearing as the forms of ${\cal L}_{4}$
and ${\cal L}_{5}$ in Galileon gravity \cite{Nicolis,DEV}.

\subsection{Brans-Dicke theories with a universal dilatonic coupling}
\label{mosec2} 

The Brans-Dicke (BD) theory \cite{Brans} is characterized
by the action 
\begin{equation}
S=\int d^{4}x\sqrt{-g}\left[\frac{1}{2}\chi R
-\frac{\omega_{{\rm BD}}}{2\chi}(\nabla\chi)^{2}+\cdots\right]\,,\label{BDaction}
\end{equation}
where $\chi$ is the scalar field coupled to $R$, and $\omega_{{\rm BD}}$
is the BD parameter. Introducing the field $\phi$ in the form 
$\chi=M_{{\rm pl}}^{2}e^{-2Q\phi/M_{{\rm pl}}}$,
the action (\ref{BDaction}) can be written as \cite{Yoko} 
\begin{equation}
S=\int d^{4}x\sqrt{-g}\left[\frac{1}{2}F(\phi)R+
(1-6Q^{2})\frac{F(\phi)}{M_{{\rm pl}}^{2}}X+\cdots\right]\,,\label{BDaction2}
\end{equation}
 where 
\begin{equation}
F(\phi)=M_{{\rm pl}}^{2}e^{-2Q\phi/M_{{\rm pl}}}\,,\qquad X=-\frac{1}{2}(\nabla\phi)^{2}\,,\qquad Q^{2}=\frac{1}{2(3+2\omega_{{\rm BD}})}\,.\label{omeq}
\end{equation}
If we define the field $\varphi$ as $\varphi=2Q\phi$,
the square bracket in Eq.~(\ref{BDaction2}) is expressed 
as ${\cal L}=M_{{\rm pl}}^{2}e^{-\varphi/M_{{\rm pl}}}[R-\omega_{{\rm BD}}(\nabla\varphi)^{2}]/2+\cdots$.
This means that dilaton gravity \cite{Gas} corresponds to 
$\omega_{{\rm BD}}=-1$.
Motivated by dilaton gravity, we shall consider the theories in which
the field $\phi$ has a universal coupling 
$F(\phi)=M_{{\rm pl}}^{2}e^{-2Q\phi/M_{{\rm pl}}}$
with $X\square\phi$ as well: 
\begin{equation}
S=\int d^{4}x\sqrt{-g}\left[\frac{1}{2}F(\phi)R+(1-6Q^{2})
\frac{F(\phi)}{M_{{\rm pl}}^{2}}X-\frac{\lambda F(\phi)}
{M^{3}M_{{\rm pl}}^{2}}X\square\phi\right]\,,
\label{BDaction3}
\end{equation}
where $\lambda$ is a constant of the order of unity.
The last term appears as the $\alpha'$ correction in low-energy
effective string theory \cite{alpha}. 
We assume that the coupling $Q$ is of the order of unity 
with $Q^2 \neq 1/6$.
The action (\ref{BDaction3}) corresponds to the theories with
 $f(\phi)=(1-6Q^{2})F(\phi)/M_{{\rm pl}}^{2}$,
$g(\phi)=\lambda F(\phi)/M_{{\rm pl}}^{2}$, and $n=1$ in Eq.~(\ref{modelcon}).
We consider the case in which the field satisfies the boundary condition 
$|\phi(0)| \ll M_{\rm pl}$ at the origin. 
Since the field derivative $\phi'(r)$ is small, 
the condition $|\phi(r)| \ll M_{\rm pl}$ is satisfied for $r>0$.
{}From Eq.~(\ref{omeq}) dilaton gravity ($\omega_{{\rm BD}}=-1$)
corresponds to $Q^{2}=1/2$.

{}From Eq.~(\ref{rvdef}) the Vainshtein radius is given by 
\begin{equation}
r_{V}\approx\left(\frac{|4Q \lambda|}{(1-6Q^{2})^{2}}
\frac{M_{{\rm pl}}r_{g}}{M^{3}}\right)^{1/3}\,,\label{rvbd}
\end{equation}
 where we used the approximations $F\approx M_{{\rm pl}}^{2}$ and
$B\approx1-6Q^{2}$.
If the model (\ref{BDaction3})
is responsible for the late-time cosmic acceleration, one can show
that there is a de Sitter solution characterized by $\dot{\phi}={\rm constant}$
with $M^{3}\approx M_{{\rm pl}}r_{c}^{-2}$. For $|Q|$ and $|\lambda|$
of the order of unity the Vainshtein radius (\ref{rvbd}) is estimated as
$r_{V}\approx(r_{g}r_{c}^{2})^{1/3}$.

Under the condition $|Q\rho_c \lambda| \gg M_{\rm pl}M^3 (1-6Q^2)^2$,
the solution in the regime $r\ll r_{*}$ is given by 
$\phi'(r)\simeq-[|Q|M^{3}\rho_{c}/(6M_{{\rm pl}}\lambda)]^{1/2}r$ for
$Q^2<1/6$ and $\phi'(r)\simeq[|Q|M^{3}\rho_{c}/
(6M_{{\rm pl}}\lambda)]^{1/2}r$ for $Q^{2}>1/6$.
The matching radius $r_{*}$ given in Eq.~(\ref{rstar})
can be estimated as
\begin{equation}
\frac{r_{*}}{r_{g}} 
\approx
\left[\frac{2|Q \lambda|}
{(1-6Q^{2})^{2}}\frac{r_{V}}{r_{g}}\frac{\rho_{0}}{\rho_{c}}
\left(\frac{r_{c}}{r_{V}}\right)^{4}\right]^{1/3}\,,
\end{equation}
where we used the relations 
$M^{3}\approx M_{{\rm pl}}r_{c}^{-2}$
and $\rho_{0}\approx3M_{{\rm pl}}^{2}/r_{c}^{2}$. 
The radius $r_*$ is the same order as Eq.~(\ref{rr}) 
for $|Q \lambda|={\cal O}(1)$, 
so that the solutions to the field equation derived in Sec.~\ref{vasec}
are again trustable.

Equations (\ref{Phiso}) and (\ref{Psiso}) show that, in the regime
$r_{*}\ll r\ll r_{V}$, the gravitational potentials can be 
estimated as 
\begin{equation}
\Phi\simeq\frac{r_{g}}{2r}\left[1-\frac{2Q^{2}}{1-6Q^{2}}\left(\frac{r}{r_{V}}\right)^{3/2}\right]\,,\qquad\Psi\simeq-\frac{r_{g}}{2r}\left[1-\frac{4Q^{2}}{1-6Q^{2}}\left(\frac{r}{r_{V}}\right)^{3/2}\right]\,.
\end{equation}
The experimental bound (\ref{bound}) translates into 
\begin{equation}
\left(\frac{r}{r_{V}}\right)^{3/2}<2.3\times10^{-5}\frac{|1-6Q^{2}|}{2Q^{2}}\,.
\end{equation}
In dilaton gravity ($Q^2=1/2$), for example, this constraint is 
satisfied for $r<10^{-3}r_{V}\approx10^{17}$\,cm for the Sun.
This upper bound is much larger than the solar-system scales.

\section{Conclusions}
\label{consec} 

In this paper we have studied the Vainshtein mechanism in second-order
scalar-tensor theories given by the action (\ref{action}). In a spherically
symmetric background the full equations of motion were derived in
the presence of a barotropic perfect fluid. Introducing the small
parameters $\epsilon_{i}$ defined in Eq.~(\ref{epori}) and picking up
the dominant contributions in the weak gravitational background, we
obtained the closed-form equations for the field $\phi$ and for the
gravitational potential $\Psi$. The approximation employed in Sec.~\ref{fieldeqsec}
is valid under the conditions $|\epsilon_{i}|\ll1$, which is required
for the consistency with solar-system experiments.

The general theories in which the Vainshtein mechanism can be at work
are given by the action (\ref{modelcon}) with the nonminimal coupling
$F(\phi)=M_{\rm pl}^2 e^{-2Q\phi/M_{\rm pl}}$. 
This action covers a wide range of modified gravitational theories such
as the (extended) Galileon, dilaton gravity, and Brans-Dicke theories
with the nonlinear field interaction. In such theories we derived
the general formula (\ref{rvdef}) for the Vainshtein radius $r_{V}$.

For $Q \neq 0$ the solution to the field equation in the regime 
$r\gg r_{V}$ is given by Eq.~(\ref{phipgene}),
which leads to the large modification to the gravitational potentials.
In the regime $r_{*}\ll r\ll r_{V}$ the solution
changes to Eq.~(\ref{phiva}), so that the modification of gravity
is suppressed even for $|Q|={\cal O}(1)$. In this regime we derived
the analytic solutions for the gravitational potentials and showed
that the experimental bound on the post-Newtonian parameter $\gamma$
can be satisfied under the condition (\ref{bound}).
In the regime $r\ll r_{*}$ the solution to the field equation for
$Q\neq0$ is given by $\phi'(r)\propto r$, which satisfies 
the regularity condition $\phi'(0)=0$ at the center of the spherical
symmetry. 

When $Q=0$ and $n=1$, if the same model is responsible for the 
late-time cosmic acceleration, the corrections to the gravitational 
potentials are extremely tiny
on solar-system scales. For $Q=0$ and $n>1$ we showed that the solution
around the origin is not regular and hence this theory cannot 
be regarded as a viable one.

We applied our general results to concrete theories such as extended
Galileon and Brans-Dicke theory with a dilatonic coupling. For the
theories with $Q\neq0$ and $n=1$ there is a correction of the order
of $(r/r_{V})^{3/2}$ relative to the Newtonian gravitational potentials
in the regime $r_{*}\ll r\ll r_{V}$,
such that the local gravity constraints can be satisfied on solar-system
scales. Note that for $Q=0$ and $n=1$ the corrections to the gravitational
potentials are even much smaller. If the extended Galileon theory
with $Q\neq0$ and $n>1$ is responsible for the cosmic acceleration
today, we found that there is a problem of the matching at $r=r_{*}$.

It will be of interest to see how the Vainshtein mechanism works in
the Horndeski's most general scalar-tensor theories having the term
$G_{5}$ as well as the $X$-dependence in $G_{4}$. 
The construction of viable dark energy models 
satisfying recent experimental and observational bounds
will be also interesting.
In particular the constraint coming from the variation of the Newton constant ($|\dot{G}/G|<0.02H_0$) can provide tight bounds on such models \cite{BabiPRL}. 
We leave these issues for future work.

\section*{ACKNOWLEDGEMENTS}
\label{acknow} The work of A.\,D.\,F.\ and S.\,T.\  was supported by the
Grant-in-Aid for Scientific Research Fund of the JSPS Nos.~10271
and 30318802. S.\,T.\ also thanks financial support for the Grant-in-Aid
for Scientific Research on Innovative Areas (No.~21111006). 
We thank M.~Sami for his initial collaboration on this work.


\end{document}